\documentclass[aps,prl,preprint,groupedaddress]{revtex4-1}

\def\be{\begin{equation}}
\def\ee{\end{equation}}
\def\beq{\begin{eqnarray}}
\def\eeq{\end{eqnarray}}
\def\n{\nonumber}

\usepackage{graphicx}

\begin{document}


\title{Inverse square law isothermal property in relativistic charged static distributions }


\author{Sudan Hansraj}

\altaffiliation{}
\affiliation{Astrophysics and Cosmology Research Unit, University of KwaZulu Natal}
\email[]{hansrajs@ukzn.ac.za}
\author{Nkululeko Qwabe}

\altaffiliation{}
\affiliation{ Astrophysics and Cosmology Research Unit, University of KwaZulu Natal}
\email[]{nkuluq@gmail.com}

\date{\today}

\begin{abstract}

We analyse the impact of the inverse square law fall-off of the energy density in a charged isotropic spherically symmetric fluid. Initially we impose a linear barotropic equation of state $p=\alpha \rho$ but this leads to an intractable differential equation. Next we consider the neutral isothermal metric of Saslaw, Maharaj and Dadhich (1996) in an electric field and the usual inverse square law of energy density and pressure results thus preserving the equation of state. Additionally, we discard a linear equation of state and endeavour to find new classes of solutions with the inverse square law fall off of density. Certain prescribed forms of the spatial and temporal gravitational forms result in new exact solutions. An interesting result that emerges is that while isothermal fluid spheres are unbounded in the neutral case, this is not so when charge is involved. Indeed it was found that barotropic equations of state exist and hypersurfaces of vanishing pressure exist establishing a boundary in practically all models.   One model was studied in depth and found to satisfy other elementary requirements for physical admissability such as a subluminal sound speed as well as gravitational surface redshifts smaller than 2. The Buchdahl (1959), Bohmer and Harko (2007) and Andreasson (2009) mass-radius bounds were also found to be satisfied.   Graphical plots utilising constants selected from the boundary conditions established that the model displayed characteristics consistent with physically viable models.

\end{abstract}


\maketitle

\section{Introduction}

In isothermal  models of the universe, the  metrics  have the property that pressure gradients  balance the mutual self-gravity of their own constituent particles. In this scenario  galaxies are considered to be idealized points due to the scale.  The movement of the particles and their velocity is independent of  their position.  Consequently these particles appear to obey the equation of state $ p= \alpha \rho$, where $p$ is the pressure and $\rho$ is the matter density.  The equation of state gives the isotropic particle pressure as a function of the density $p=p(\rho)$ and expresses the fact that the temperature is independent of the position within the distribution.  In other words the pressure is proportional to the density irrespective of the location within the sphere. The parameter $\alpha$ is a constant satisfying $0<\alpha\leq 1$ in order to ensure that the fluid remains causal, that is the speed of sound never exceeds the speed of light.   This model then results in a configuration that is neither expanding nor contracting and  which is understood to be a global solution that is stationary.  In cosmological cases, the particle motion is taken to be non-relativistic.  In the  Newtonian analogue  $\rho$ is finite at the core but decreases as $r^{-2}$ throughout most of the configuration by the prescribed equation of state $p=\alpha \rho$ and so it follows that $p \propto r^{-2}$ .  This then implies that isothermal fluids are by design unbounded, there being no possibility of a surface of vanishing pressure.    Hence the total mass and the radius of the isothermal  sphere is infinite. Moreover the prescription $\rho \propto r^{-2}$ throughout the entire sphere results in the density of the sphere having a singularity at the center.  Consequently the point in the isothermal model with the highest density is at the centre of the sphere.

We analyse the case of the isothermal fluid sphere with an inherent charge to investigate the role of charge in such distributions.  While the prevailing viewpoint is that astrophysical fluids are in general neutral,  it is worth considering the charged case on account of the fact that the "no-hair theorem" by Hawking and Penrose \cite{hawk} on black hole dynamics assert that the evolution of a black hole is solely dependent on its mass, charge and angular momentum. Moreover Cherubini {\it et al} \cite{cheru} have shown that charge may also  play a role in gamma ray bursts. Electromagnetic black holes were discussed in detail by Ruffini {\it et al} \cite{ruff1,ruff2,ruff3,ruff4} in the context of gamma ray bursts. Vacuum polarisation around electromagnetic black holes were studied in \cite{cheru2}. Therefore the presence of charge may not be ruled out in the phase transitions of stellar evolutions.   In this regard a number of charged sphere models have been reported  during the last century. The exterior electric field ($E$) is also characterised by an inverse square law $ E \propto \frac{1}{r^2}$ fall off as is the case for the isothermal density.

The line element for static spherically symmetric spacetimes,
in coordinates $(x^a) = (t,r,\theta,\phi)$, is taken as
\be
ds^2=-e^{2\nu(r)}dt^2 + e^{2\lambda(r)}dr^2 + r^2(d\theta^2 + \sin^2\theta d\phi^2) \label{112}
\ee
where the gravitational potentials $\nu$ and $\lambda$ are functions
only of   the spacetime coordinate $r$.
The Einstein--Maxwell equations determine the gravitational behaviour of charged fluids  and may be expressed as the system
\beq
\left[r(1-e^{-2\lambda})\right]' &=& r^2\rho+\frac{1}{2}r^2E^2  \label{2A} \\ \n \\
-(1-e^{-2\lambda})+2\nu're^{-2\lambda} &=& pr^2-\frac{1}{2}r^2E^2 \label{2B} \\ \n \\
e^{-2\lambda}\left[r(\nu' - \lambda') +r^2(\nu''-\nu'\lambda'+\nu'^2)\right] &=& pr^2+\frac{1}{2}r^2E^2 \label{2C} \\ \n \\
\sigma^2 &=& \frac{e^{-2\lambda}}{r^2}(r^2E)' \label{2D}
\eeq
for the static spherically symmetric spacetime (\ref{112}) and where $'$ is $\frac{d}{dr}$. The detailed derivation of this system of equations may be found in \cite{hans1,hans2,hans3}.
The conservation laws $T^{ab}_{}{}_{;b} = 0 $ reduce to the equation $
p' + (\rho + p)\nu' = \frac{E}{r^2}\left[r^2 E\right]' $ which can substitute one of the field equations in the system (\ref{2A}) to (\ref{2D}).  Defining the mass $m$ of a spherical distribution of perfect fluid within a radius $R$ by $m(R)=\int_0^R \rho  r^2 \, dr$ the conservation
 equation  yield the Tolman--Oppenheimer--Volkoff (TOV) equation or equation of hydrostatic equilibrium and  which has been used extensively to find exact solutions historically.
The exterior gravitational field for a charged fluid sphere is given by the  Reissner--Nordstr\"{o}m \cite{reis,nord} metric
 \be
ds^2=-\left( 1-\frac{2m(r)}{r} +\frac{q^2}{r^2}\right)dt^2 + \left( 1-\frac{2m(r)}{r}+\frac{q^2}{r^2} \right)^{-1} dr^2 +
r^2(d\theta^2 + \sin^2 \theta d\phi^2) \label{532xy}
\ee
where $q$ is the charge component and $m$ the mass.  It is required that both gravitational potentials and the charge across the boundary are continuous.  The Israel--Darmois junction conditions also include that $
 e^{2\nu(R)}= e^{-2\lambda(R)} = \left( 1-\frac{2M}{R} +\frac{Q^2}{R^2}\right) $ and $p(R) =0$. In addition the electric field external to a charged sphere has the form $E=\frac{Q}{R^2}$ where $r = R$ is the radius of the distribution and $Q$ is the charge measured by an  observer at spatial infinity.

  It is generally agreed (Knutsen \cite{knut}, Buchdahl \cite{buc}, Delgaty and Lake \cite{delg}) that the following conditions should be satisfied for physical reasonableness.
 The metric potentials should be free from singularities inside the radius of the star.  The pressure and density should be positive and finite inside the fluid configuration.  The  requirement imposed by some  that the pressure and density should decrease monotonically ie. $\frac{dp}{dr}<0$,  $\frac{d\rho}{dr}<0$ has been often considered excessively restrictive given that the thermodynamical processes within a star are unknown.
Causality demands that the speed of sound should never exceed the speed of light within the stellar distribution. This amounts to the condition $0< \frac{dp}{d\rho}\leq 1$.
The energy content obeys the following constraints: weak energy condition ($\rho - p > 0$), strong energy condition ($\rho + p > 0)$ and the dominant energy condition ($\rho + 3p > 0$).

In addition the gravitational surface redshift $z$ should be monotonically decreasing towards the boundary of the sphere and   the central redshift $z_0$ and the surface red shift $z_R$ should be positive and finite (Buchdahl \cite{buc}, Ivanov \cite{ivan}): $
z_c=\sqrt{e^{-\nu(0)}}-1>0 $ and $ z_R=\sqrt{e^{-\nu(R)}}-1>0 $
and in general $z<2$ is expected for relativistic stars.
 The maximum mass to radius ratio for a static fluid sphere must satisfy the
  condition $ \frac{\rm mass}{\rm radius} < \frac{4}{9}$ to ensure the stability of the sphere (Buchdahl \cite{buc}).  This means that there is a limit to the amount of matter that can be packed into a given sphere of radius $R$.
 B$\ddot{o}$hmer and Harko \cite{boh} established the following upper bound for the mass--radius--charge ratio: $\frac{Q^2(18 R^2+Q^2)}{R^2(12 R^2+Q^2)}\leq \frac{2 M}{R} $ while  Andr$\acute{e}$asson \cite{andr} determined the  lower bound
$\frac{ M}{R}\leq \left(\frac{R+\sqrt{R^2+3 Q^2}}{3 R}\right)^2$.

A literature survey reveals that many solutions reported thus far are singular at the centre and are valid only for restricted
regions of spacetime. Such solutions may be considered as core--envelope models \cite{hans4} where the core consists of a different material, possibly uncharged. The two metrics will need to be matched at the common interface.  Herrera and Ponce de Leon \cite{herr}, Pant and Sah \cite{pant}, Tikekar \cite{tik} and Whitman and Burch \cite{whit} generated models with a  singularity at the  stellar centre.   Maartens and Maharaj \cite{mar} presented a solution that was stable and regular at the center but  their solution had negative pressure.  Models with vanishing pressure  were analysed by Bonnor \cite{bonn3}, Bonnor and Wickramasuriya \cite{bonn1} and Raychaudhuri \cite{ray5}.     De and Raychaudhari \cite{de} have verified  that in order to guarantee the equilibrium of a static charged dust sphere the relation $\sigma=\pm\rho$ must be satisfied.   The general forms for charged de Sitter solutions, the energy--momentum tensor with a constant energy component, pressure including charged dust and linear equation of state were reviewed by Ivanov \cite{ivan}.

\section{Isothermal fluid field equations}

Implementing the coordinate transformations $x=Cr^2$ for some constant $C$, $Z(x)= e^{-2\lambda(r)}$ and  $y^2(x)= e^{2\nu(r)}$
 the Einstein--Maxwell field equations (\ref{2A}) -- (\ref{2D}) assume the form
\beq \label{2aa}
\frac{1-Z}{x}-2\dot{Z} &=& \frac{\rho}{C}+\frac{E^2}{2C}  \label{1} \\ \n \\
\frac{Z-1}{x} + \frac{4Z\dot{y}}{y} &=& \frac{p}{C}-\frac{E^2}{2C} \label{2} \\ \n \\
4x^2Z\ddot{y} + 2x^2\dot{Z}\dot{y}+\left(\dot{Z}x-Z+1-\frac{E^2x}{C}\right)y &=& 0 \label{3} \\ \n \\
\frac{\sigma^2}{C} &=& \frac{4Z}{x}(x\dot{E}+E)^2 \label{4}
\eeq
The benefit of the change of coordinates introduced is that the equation of pressure isotropy (\ref{3}) is linear second order in the variable $y$ whereas it is nonlinear first order in $Z$.  The field equation (\ref{3}) may be viewed as the master equation for this system. Once a form for $Z$ is chosen, we may proceed with the possible integration of (\ref{3}). A large number of exact solutions have been discovered in this manner. For example see Thirukannesh and Maharaj \cite{thi}, Finch and Skea \cite{fin}, Maharaj and Mkhwanazi \cite{mah3}. The last named authors actually regained the Schwarzschild interior solution and demonstrated the equivalence of their solution with the Schwarzschild interior solution.  Note that the dynamic quantities may be written in terms of the metric potentials in the following way:
\beq
\frac{\rho}{C} &=& \frac{-4x^2Z\ddot{y}-2x^2\dot{Z}\dot{y}+(1-Z-5x\dot{Z})y}{2xy}  \label{22A} \\ \n \\
\frac{p}{C} &=& \frac{4x^2Z\ddot{y}+2x(4Z+x\dot{Z})\dot{y}+(Z-1+x\dot{Z})y}{2xy} \label{22B} \\ \n \\
\frac{E^2}{C} &=& \frac{4x^2Z\ddot{y}+2x^2\dot{Z}\dot{y}+(x\dot{Z}-Z+1)y}{xy} \label{22C} \\ \n \\
\frac{\sigma^2}{C} &=& \frac{4Z}{x}\left( \frac{d}{dx}(xE)\right )^2 \label{22D}
\eeq
From this presentation of the field equations it is patently clear that any metric with components $\nu$ and $\lambda$  will be suitable for generating a complete model without having to perform any integrations.  In other words this demonstrates that devising charged star models is a trivial exercise. See for example the treatment of Krori and Barua \cite{krori}.  The caveat in this approach however, is that by prescribing the metric, all control over the physics of the problem is relinquished.  The possibility that the fluid may display an equation of state is now very remote although not impossible.

Let us examine the problem of imposing the inverse square law on the density as well as the linear barotropic equation of state $p=\alpha \rho$. These two conditions should generate a unique solution as the system is now sufficiently determined.
Examining the field equations we observe that  (\ref{1}) and (\ref{2}) yields
\begin {equation}
-2\dot{Z}+\frac{4Z\dot{y}}{y}=\frac{\rho+p}{C}\label{5}
\end{equation}
 where the quantity $\displaystyle{{\rho+p}}$ is referred to as the inertial gravitational  energy density.   Next we invoke the isothermal conditions $\rho \sim \frac{1}{r^2}$ and $p =\alpha \rho$ as per Saslaw {\it et al} \cite{sas}.  Let ${\rho=\frac{A}{x}}$ and ${p=\frac{B}{x}}$ where $A$ and $B$ are suitable constants. Then  (\ref{5}) takes the form
\begin {equation}
-2\dot{Z}+\frac{4Z\dot{y}}{y}=\frac{A+B}{Cx}\label{6}
\end{equation}
which is linear first order in both $Z$ and $y$.  We may rearrange (\ref{6}) in the form
\begin {equation}
\frac{\dot{y}}{y}=\frac{(A+B)+2\dot{Z}xC}{4ZxC}\label{8}
\end{equation}
in order to separate the variables.
 Differentiating equation (\ref{8}) with respect to $x$ we obtain the relationship
\begin{equation}
\frac{\ddot{y}}{y}-\left(\frac{\dot{y}}{y}\right)^2=-\frac{A+B+2 C x \dot{Z}(x)}{4 C x^2 Z(x)}-\frac{\dot{Z}(x) \left(A+B+2 C x \dot{Z}(x)\right)}{4 C x Z(x)^2}+\frac{2 C x \ddot{Z}(x)+2 C \dot{Z}(x)}{4 C x Z(x)}\label{9}
\end{equation}
relating $\ddot{y}$ to $\dot{y}$  and $y$.  Inserting (\ref{9}) into (\ref{3}) we generate the equation
\beq
{2\dot{Z}x+2x^2\ddot{Z}}-\frac{\left(A+B+2CX\dot{Z}\right)(4CZ+4xC\dot{Z})}{4C^2Z}+\left[\frac{(A+B)+2Cx\dot{Z}}{4xCZ}\right]^2 && \n \\ \n \\  + xZ\left[\frac{(A+B)+2CZ\dot{Z}}{2CZ}\right]+\dot{Z}x-Z+1-\frac{E^2}{C}x &=&0 \label{10qq}
\eeq
containing  only $Z$ and $E$ .   Using the expression for $\displaystyle{\frac{E^2}{C}}$ from (\ref{1}),  equation (\ref{10qq}) assumes the form
\begin{equation}
8C^2x^2Z \ddot{Z}+4C^2Z \dot{Z}(1+4x)+2Cx\dot{Z}(A+B)+4C^2Z^2+4CZ(A-B-C)+(A+B)^2=0\label{10}
\end{equation}
which is a complicated nonlinear second order ordinary differential equation.  Solving (\ref{10}) has proven to be an intractable problem.  This illustrates the difficulty of imposing an equation of state on static isotropic models early. This is a simple model yet it has floundered on account of the severe nonlinearity of the master isotropy equation.  This difficulty was discussed very soon after the advent of general relativity by  Tolman \cite{tol} who produced non--trivial new exact solutions for neutral fluids some twenty years after the publication of the Schwarzschild solutions \cite{schwar1,schwar2}.

\section{Saslaw, Maharaj and Dadhich model in an electric field}

Saslaw {\it et al}  \cite{sas} have investigated  the isothermal sphere for neutral compact stars.  Observe that the problem of solving the Einstein field equations for fluid spheres amounts to solving a system of three partial differential equations in four unknowns $\rho, p,\nu, \lambda$ in the uncharged case.  Therefore  specifying any ONE of these quantities will close the system.  Then a unique solution exists theoretically.  However  Saslaw {\it et al} in specifying isothermal behavior   $\left({\rho \sim \frac{1}{r^2} }; {p \sim \frac{1}{r^2}}\right)$ in effect are determining two of the variables upfront.  This ordinarily means that the system is over determined.  But in their treatment, the equation of pressure isotropy is utilised  as a consistency condition and has the effect of determining unknown constants.  They have demonstrated adequately  all the field equations are indeed satisfied.  It still is an open question whether the metric reported is the most general.  Hansraj {\it et al} \cite{hans5} have analysed this question in the framework of free--trace Einstein gravity and have produced more general behaviour than that found by Saslaw {\it{et al}}.

The Saslaw {\it {et al}} metric potentials expressed $e^\nu =\xi r^{4\alpha/(1+\alpha)}$ and $ e^\lambda =1+\frac{4\alpha}{(1+\alpha)^2}$
are inserted into equations  (\ref{22A}) -- (\ref{22D}).  Note that the gravitational potential $\lambda$ is constant.  Dadhich {\it{et al}} \cite{ dad1}  have shown that a constant gravitational potential is a necessary and sufficient  requirement for isothermal behaviour not only in the Einstein framework of general relativity but also in the more general theory of pure Lovelock gravity which contains the Einstein case for the first order.   However it must be noted that in the present case of a charged sphere the equation of state $p=\alpha \rho$ is not guaranteed.  It remains to be seen whether a barotropic equation of state exists in the presence of the electric field.

Rewriting the Saslaw {\it{et al}} potentials in our transformed coordinate system with $x=Cr^2$, we obtain
\be
y=e^{2\nu}=\xi\left(\frac{x}{C}\right)^{\frac{2\alpha}{1+\alpha} } {\hspace{1cm}}
Z=e^{-2\lambda}=\left(\frac{(1+\alpha)^2}{(\alpha^2+6\alpha+1)}\right)^2\label{13c}
\ee
where $\displaystyle{\xi}$ is a constant.
 When these metric components are substituted  into (\ref{22A}) to (\ref{22D}) we get
\beq
\rho &=&-\frac{4 \alpha  \left(\alpha ^3-5 \alpha -2\right)}{\left(\alpha ^2+6 \alpha +1\right)^2 r^2} \label{22AA} \\ \n \\
p &=& \frac{4 \alpha ^2 \left(3 \alpha ^2+6 \alpha +1\right)}{\left(\alpha ^2+6 \alpha +1\right)^2 r^2} \label{22BB} \\ \n \\
E^2 &=&\frac{8 \alpha ^2 \left(\alpha ^2+2 \alpha +3\right)}{\left(\alpha ^2+6 \alpha +1\right)^2 r^2} \label{22CC} \\ \n \\
\sigma^2 &=&\frac{8 \alpha ^2 (\alpha +1)^4 \left(\alpha ^2+2 \alpha +3\right)}{\left(\alpha ^2+6 \alpha +1\right)^4  r^4} \label{22DD}
\eeq
for the dynamical and electric  quantities. Observe that the equation of state ${p=\alpha \rho}$ is maintained and both density and pressure are inversely proportional to ${r^2}$.  That is the isothermal behavior is preserved despite the introduction of charge.  In fact the charge density $\sigma$ also obeys the inverse square law while the electric field intensity varies according to $E \propto \frac{1}{r}$ in the interior in contrast with the exterior where the inverse square law applies.
The expressions that define the energy conditions are expressed as follows:
\beq
{p}+{\rho}&=&\frac{8 \alpha  (\alpha +1)^3}{\left(\alpha ^2+6 \alpha +1\right)^2 r^2}\label{26d}\\ \n \\
{\rho}-{p}&=&-\frac{8 \alpha  \left(2 \alpha ^3+3 \alpha ^2-2 \alpha -1\right)}{\left(\alpha ^2+6 \alpha +1\right)^2 r^2}\label{26e}\\ \n \\
{\rho}+{3p}&=&\frac{8 \alpha  \left(4 \alpha ^3+9 \alpha ^2+4 \alpha +1\right)}{\left(\alpha ^2+6 \alpha +1\right)^2 r^2}\label{26f}
\eeq
for the weak, strong and dominant  energy conditions respectively.

In order to guarantee a subluminal sound speed, it is required  that ${0<\frac{dp}{d\rho}<1}$. This translates to the relationship
$ 0<\frac{\alpha  \left(3 \alpha ^2+6 \alpha +1\right)}{-\alpha ^3+5 \alpha +2}<1 $
constraining $\alpha$.  Calculating the  acceptable range results  in ${0<\alpha< 0.744644}$  after discarding options with a negative $\alpha$ to avoid a violation of causality.
  To ensure a positive energy density requires
$0<\alpha <1+\sqrt{2}$. Likewise for a positive pressure, any $\alpha > 0$ is suitable. This is also true to ensure the positivity of $E^2$, $\sigma^2$, $\rho + p$ and $\rho + 3p$. The weak energy condition $\rho > p$ demands that $0< \alpha < 0.744644$. In summary all the physical requirements are satisfied for $0 < \alpha < 0,744644$.
  This demonstrates that  the Saslaw {\it et al} metric does admit a model of an isothermal relativistic fluid incorporating charge that is well behaved away from the centre of the distribution.

\section{Charged spheres with inverse square fall-off of density }

In the previous section we analysed the isothermal condition $\rho \propto \frac{1}{r^2}$ and $p=\frac{1}{r^2}$ which implied that $p \propto \rho$.    What is important about this system is that the only factor affecting the pressure and hence the density is the velocity (momentum) of the electrons and the change in the temperature has no influence in this system hence this condition is called the isothermal equation of state.
It is now interesting to ask what solutions are admitted for the isothermal prescription if we abandon the linear equation of state.  Of course other equations of state may be possible.  That is we prescribe $\rho \propto \frac{1}{r^2}$ but allow the pressure to take on a variety of profiles.  The problem then still admits an infinite number of solutions in theory as one quantity remains to be specified. In this section we study various choices for $Z$ and $y$  in order to fully integrate the Einstein--Maxwell field equations   (\ref{1}) to (\ref{4}) subject to the density varying according to the inverse square law.

Consider the transformed Einstein--Maxwell field equations  (\ref{1}) to (\ref{4}).  Introducing the condition $\rho=\frac{k}{x}$ for some constant $k$  equation (\ref{1}) reduces to
\begin {equation}
\frac{E^2}{C}=\frac{2(1-Z)}{x}-4 \dot{Z}-\frac{2k}{x} \label{26a}
\end{equation}
 relating the electric field $E$ to $Z$.  Then substituting (\ref{26a}) into (\ref{3}) results in a second order differential equation
\begin{equation}
4x^2Z\ddot{y}+2x^2 \dot{Z} \dot{y}+\left(5\dot{Z}x+Z-1+2k\right)y=0 \label{26q}
\end{equation}
expressing the relationship between the metric potentials $Z$ and $y$.  Equation(\ref{26q}) is the master equation for Einstein--Maxwell perfect fluids with the density obeying the inverse square law.  In order to close the system, functional forms for one of $Z$ or $y$ may be chosen to integrate (\ref{26q}).

\subsection{Specifying the spatial gravitational potential}

\begin{itemize}
\item{The case $Z=$ a constant}

Though this is the simplest form of $Z$, it is a nontrivial choice for the gravitational potential.   If $Z$ is taken to be a constant $\beta$ in (\ref{26q}),  the following form of the remaining potential
\begin{equation}
y= x^{\frac{1}{2}-\frac{\sqrt{1-2 k}}{2 \sqrt{\beta }}} \left(c_2 x^{\frac{\sqrt{1-2 k}}{\sqrt{\beta }}}+c_1\right) \label{26q1}
\end{equation}
results. Here $c_1$ and $c_2$ are constants of integration.  The complete model is now given by
\beq
\frac{\rho}{C} &=& \frac{k}{x} \label{26q1a} \\ \n\\
{\frac{p}{C}}&=& \frac{ 2 \beta  \left(c_2 x^{\gamma}+c_1\right) -k \left(c_2 x^{\gamma}+c_1\right) -2 \sqrt{\beta } \sqrt{1-2 k} \left(c_1-c_2 x^{\gamma}\right)}{x \left(c_2 x^{\gamma}+c_1\right)} \label{26q1b} \\ \n\\
{\frac{E^2}{C}}&=&\frac{2 (1-k-\beta )}{x} \label{26q1a} \\ \n\\
\frac{\sigma^2}{C}&=&\frac{2 \beta  C^2 (1-k-\beta)}{x^2}\label{26q1c}
\eeq
where we have set ${\gamma=\sqrt{\frac{1-2 k}{ \beta}}}$ for simplicity.
Observe that the constant potential does not produce exactly isothermal behaviour in the presence of charge.  The functional form of the pressure is no longer of the form ${\frac{1}{r^2}}$.  However an equation of state does exists.  From (\ref{26q1a}) we get ${x=\frac{k C}{\rho}}$ and inserting into   (\ref{26q1b}) gives
\begin{equation}
p=\rho  \left(\frac{4 \sqrt{\beta } c_1 \sqrt{1-2 k}}{c_2 k \left(-C k\rho^{-1} \right)^{\gamma}+c_1 k}-\frac{2 \left(\beta +\sqrt{\beta } \sqrt{1-2 k}\right)}{k}+1\right) \label{26q0}
\end{equation}
which is a  barotropic equation of state.  Note that when $k= \beta-1$, the electric field $E$ and the proper charge density vanish so that the fluid is now neutral.  It is also evident that a surface of zero pressure $p(R)=0$ exists at
\[x=CR^2=\left(\frac{-4 \beta  c_1(\beta +1) -c_1 k^2   +12 \beta  c_1 k +4  c_1 \sqrt{\beta(1-2 k)}(2\beta -  k)}{4 \beta ^2 c_2-4 \beta  c_2+c_2 k^2+4 \beta  c_2 k}\right)^{{\frac{1}{\gamma}}} \]
The speed of sound index is given by 
\begin{equation}
\frac{dp}{d\rho}= \frac{4 c_1 c_2 x^{\gamma}-2 \beta  \left(c_2 x^{\gamma}+c_1\right){}^2+k \left(c_1^2 -6 c_1 c_2 x^{\gamma}+c_2^2 x^{2\gamma}\right)+2  \sqrt{\beta}k_1 \left(c_1^2-c_2^2 x^{2\gamma}\right)}{k \left(c_2 x^{\gamma}+c_1\right){}^2} 
\label{110cc}
\end{equation}
where we have put $k_1 =\sqrt{2k-1}$. 

The expressions that define the  energy conditions 
\beq
\frac{\rho-p}{C}&=&\frac{2 \sqrt{\beta } C \left(c_1 \left(\sqrt{1-2 k} -\sqrt{\beta }\right)- c_2 \left(\sqrt{\beta }+\sqrt{1-2 k}\right) x^{\frac{\sqrt{1-2 k}}{\sqrt{\beta }}}\right)}{x \left(c_2 x^{\gamma}+c_1\right)} \label{26q1g} \\ \n\\
\frac{\rho+p}{C}&=&\frac{2  \left( (\beta -k)\left(c_2 x^{\gamma}+c_1\right) -\sqrt{\beta } \sqrt{1-2 k} \left(c_1-c_2 x^{\gamma}\right)\right)}{x \left(c_2 x^{\gamma}+c_1\right)} \label{26q1h} \\ \n\\
\frac{\rho+3p}{C}&=&\frac{2  \left((3\beta -2k) \left(c_2 x^{\gamma}+c_1\right) -3 \sqrt{\beta } \sqrt{1-2 k} \left(c_1-c_2 x^{\gamma}\right)\right)}{x \left(c_2 x^{\gamma}+c_1\right)} \label{26q1e}
\eeq
are useful in analysing the physical properties of the model. We undertake such an analysis for a different configuration.

\item{$Z= x^n$}

The general form $Z=x^n$ for any real number $n$ when substituted in the master equation (\ref{26q}) yields the solution
\be
y=  x^{-\frac{1}{4}n+\frac{1}{2}}\left( c_1 J\left(-\sqrt{\frac{n-24}{4n}},\frac{\sqrt{2k-1}x^{-\frac{1}{2}n}}{n}\right) +c_2 Y\left(-\sqrt{\frac{n-24}{4n}},\frac{\sqrt{2k-1}x^{-\frac{1}{2}n}}{n}\right) \right) \label{27a}
\ee
where $J$ and $Y$ are Bessel functions of the first  and second kind. It is well known that Bessel functions of half-integer order are realisable as elementary functions. Note also that to maintain a real valued order of the Bessel function requires $n<0$ or $n>24$. Additionally for half integer orders we must have $n=\frac{24}{1-u^2}$ where $u$ is an odd number. For $n < 0$ we get $u <-1$ or $u>1$ and so an infinite number of suitable odd numbers exist producing solutions in terms of elementary functions. On the other hand, the case $n>24$ is not viable as this implies $-1 < u < 1$ and there are no odd numbers in this interval.  We   list below some special cases for $n$  that generate closed form exact solutions.

\begin{itemize}
\item{} $n=-1$
\beq
y \left( x \right) ={\frac {{C_1}\, \left( \,k_1
 \left(   k_1^2  {x}^{3/2}-3\,\sqrt {x} \right) \sin
 \left( k_1\sqrt {x} \right) +\,3 \left( k-1 \right)x
\cos \left( k_1\sqrt {x} \right)  \right) }{x}}
\n \\ \n \\
+{\frac {{
C_2}\, \left( \,k_1 \left(   k_1^2  {x}
^{3/2}-3\,\sqrt {x} \right) \cos \left( k_1\sqrt {x}
 \right) -3\, \left( k-1 \right) x\sin \left( k_1\sqrt {x
} \right)  \right) }{x}}
\eeq
where we have redefined $k_1 =\sqrt{2k-1}$.

\item{} $n=-3$
\beq
y= c_1 k_1^{\frac{1}{3}} \sqrt{x} \left(2 \sinh \left(\frac{1}{3} k_1 x^{\frac{3}{2}}\right)-\frac{6 \cosh \left(\frac{1}{3} k_1 x^{\frac{3}{2}}\right)}{k_1 x^{\frac{3}{2}}}\right)
\n \\ \n \\
- c_2 k_1^{\frac{1}{3}} \sqrt{x} \left(2 \cosh \left(\frac{1}{3} k_1 x^{\frac{3}{2}}\right)-\frac{6 \sinh \left(\frac{1}{3} k_1 x^{\frac{3}{2}}\right)}{k_1 x^{\frac{3}{2}}}\right)
\eeq

\item{} $n=-\frac{1}{2}$
\beq
y  &=& c_1\, \left(  \left( -{\frac {5}{32}}\,
\sqrt {x}+\frac{1}{4} k_1^2 x \right) k_1^2 \sin
 \left( 2\,k_1\sqrt [4]{x} \right) \right.
 \n \\ \n \\   && \left.
 -\frac{1}{6}\,\cos \left( 2\,
k_1\sqrt [4]{x} \right)  k_1^3
 \left( \frac{1}{2} k_1^2 {x}^{\frac{5}{4}}-{\frac {15}{8}}\,{x}^{\frac{3}{4}}
 \right)  \right) {x}^{-\frac{3}{4}}
 \n \\ \n \\ &&
 +c_2\, \left(  \left( -{\frac {5}{
32}}\,\sqrt {x}+ \frac{1}{4} k_1^2 x  \right)  k_1^2
\cos \left( 2\,k_1\sqrt [4]{x} \right) \right.
\n \\ \n \\ && \left.
+\frac{1}{6}\, k_1^3 \left(\frac{1}{2} k_1^2 {x}^{\frac{5}{4}}-{\frac {15}{8}}
\,{x}^{\frac{3}{4}} \right) \sin \left( 2\,k_1\sqrt [4]{x} \right)
 \right) {x}^{-\frac{3}{4}}
\eeq

It is now straightforward to obtain all the physical variables for the complete model. Other functional forms for $Z$ do not appear to yield closed form solutions.

\end{itemize}

\end{itemize}

\subsection{Specifying the temporal gravitational potential}

 Equation (\ref{26q}) may be be rearranged as a differential equation in terms of $Z$ and $\dot{Z}$ in the form
\begin{equation}
(2x^2\dot{y}+5xy)\dot{Z}+(4x^2\ddot{y}+y)Z+(2k-1)y=0\label{28}
\end{equation}
which is a linear ordinary differential equation. The general solution to (\ref{28}) is given by
\be
Z=e^F\left(C + \int  \frac{e^{-F}(1-2k)y }{2x^2 \dot{y}+5xy} dx \right) \label{28a}
\ee
where $F= \int \frac{-(4x^2 \ddot{y}+y)}{2x^2 \dot{y}+5xy}dx$ and $C$ is a constant of integration.
 To establish an  exact model it now remains to stipulate functional forms for $y$ in order to integrate (\ref{28}). In theory we have located all solutions to the Einstein--Maxwell equations admitting an inverse square law fall off of the density via (\ref{28a}).

\begin{itemize}
\item{The form $y=\beta$} (Charged Einstein Universe)\\

Einstein \cite{einstein} first investigated this case and we consider it now in the context of the isothermal property.  He produced an unphysical cosmological model with a constant density and pressure.  We examine the consequences of the choice if charge is present.  With the help of (\ref{28}) we obtain
\begin{equation}
Z=\frac{c_1}{\sqrt[5]{x}}-2 k +1\label{28a}
\end{equation}
for the metric potential and where $c_1$ is a constant of integration.  The remaining dynamical and electric quantities are given by
\beq
\frac{E^2}{C} &=& \frac{2k}{x} -\frac{6c_1}{5x^{\frac{6}{5}}}  \label{28b} \\ \n \\
\frac{p}{C} &=& \frac{2 c_1}{5 x^{6/5}}-\frac{k}{x} \label{28d} \\ \n \\
\frac{\sigma^2}{C^2}&=& \frac{2  \left(25 k \sqrt[5]{x}-12 c_1\right){}^2 \left(c_1- (2 k-1) \sqrt[5]{x}\right)}{125 x^{12/5} \left(5 k \sqrt[5]{x}-3 c_1\right)} \label{28e}
\eeq
noting that a surface of vanishing pressure exists since a real root of $p(x)=0$ exists at $x=\left(\frac{2c_1}{5k} \right)^5 $.    The speed of sound index is given by
\begin{eqnarray}
\frac{dp}{d\rho} &=& \frac{12 c_1}{25 k \sqrt[5]{x}}-1 \label{28f}
\end{eqnarray}
while the functions that govern the energy conditions have the form
\beq
\frac{\rho-p}{C} &=& \frac{2 k}{x}-\frac{2 c_1}{5 x^{6/5}} \label{27a} \\ \n \\
\frac{\rho + p}{C} &=&  \frac{2 c_1 }{5 x^{6/5}} \label{27b} \\ \n\\
\frac{\rho+3p}{C} &=&  \frac{6 c_1}{5 x^{6/5}}-\frac{2 k}{x}  \label{27c}
\eeq
For the  positivity of the left hand side of (\ref{27b}) we get that $c_1 > 0$ if we take $C>0$ for $x >0$. In turn (\ref{27a}) and (\ref{27c}) results in the bound $\frac{c_1}{5k} < x^{\frac{1}{5}} < \frac{3c_1}{5k}$. From (\ref{28f}) the causality principle demands $\frac{c_1}{5k} < x^{\frac{1}{5}} < \frac{12c_1}{25k}$. Maintaining a positive pressure requires $x^{\frac{1}{5}} < \frac{2c_1}{5k}$ from  (\ref{28d}) while ensuring the positivity of $E^2$ yields $x^{\frac{1}{5}} < \frac{2c_1}{5k}$. All of these conditions constrain the radius of the sphere as $\frac{c_1^5}{3125k^5} < x < \frac{32c_1}{3125 k^5}$. In other words a physically viable model is admitted in this scenario.

There are some other interesting features inherent in this model. While the neutral version proposed by Einstein contains a constant density and pressure,  the presence of charge allows for variation of these quantities.  Moreover, the pressure is able to vanish for some value of $x$.  This suggests that a compact bounded distribution with constant gravitational potential may possess physically reasonable properties. In other words, setting the temporal gravitational potential to a constant in the neutral case generated an unphysical cosmological fluid, whereas the presence of charge admits an astrophysical fluid. Since the coordinate $x$ can always be expressed in terms of $\rho$ aa barotropic equation of state exists.  A drawback of this model is that it is difficult  to turn the electric field off.  Additionally choosing the density $\rho \propto \frac{1}{x}$ ensures a singularity at the stellar origin.  However, this singularity is irrelevant  since the electric field repulsion counters the collapse of the fluid to a central point singularity.  In other words $x=0$ is unreachable when charge is present.

\item{The form $y= 1 + x$}\\

When $y=1+x$  is substituted in (\ref{28}) the solution
\begin{equation}
Z=\frac{c_1 (7 x+5)^{2/35}}{\sqrt[5]{x}}-\frac{2 k}{C}+1\label{29}
\end{equation}
is obtained.  Observe that this metric potential contains a singularity at $x=0$.  This means that this model is viable away from the centre. The dynamical and electric quantities  are given by
\beq
\frac{E^2}{C}&=& \frac{2  \left(k \sqrt[5]{x} (7 x+5)^{\frac{33}{35}}-c_1 (5 x+3)\right)}{x^{\frac{6}{5}} (7 x+5)^{\frac{33}{35}}} \label{30b} \\ \n\\
\frac{p}{C}&=&\frac{2 \left(15 c_1 x^2+12 c_1 x+c_1+2 (7 x+5)^{\frac{33}{35}} x^{\frac{6}{5}}\right)-k \sqrt[5]{x} (7 x+5)^{\frac{33}{35}} (9 x+1)}{x^{\frac{6}{5}} (x+1) (7 x+5)^{33/35}} \label{30e} \\ \n\\
\frac{\sigma^2}{C^2}&=&\frac{2  \left(c_1 (7 x+5)^{\frac{2}{35}}+(1-2 k) \sqrt[5]{x}\right) \left(k \sqrt[5]{x} (7 x+5)^{\frac{68}{35}}-6 c_1 \left(5 x^2+7 x+2\right)\right){}^2}{x^{\frac{12}{5}} (7 x+5)^{\frac{101}{35}} \left(k \sqrt[5]{x} (7 x+5)-c_1 (5 x+3) (7 x+5)^{\frac{2}{35}}\right)} \label{30d}
\eeq
where $c_1$ is a constant of integration.  The sound speed index has the form
\beq
\frac{dp}{d\rho} &=& \left(4 \left(60 c_1 x^4+105 c_1 x^3+65 c_1 x^2+19 c_1 x+3 c_1+5 (7 x+5)^{\frac{33}{35}} x^{\frac{11}{5}}+7 (7 x+5)^{\frac{33}{35}} x^{\frac{16}{5}}\right) \right. \n \\ \n \\ && \left.
-k \sqrt[5]{x} (7 x+5)^{\frac{33}{35}} \left(63 x^3+59 x^2+17 x+5\right)\right) / \left(k \sqrt[5]{x} (x+1)^2 (7 x+5)^{\frac{68}{35}}\right) \label{30e}
\eeq
The expressions governing the energy conditions are given by
\beq
\frac{\rho-p}{C}&=&\frac{2  \left(-15 c_1 x^2-12 c_1 x-c_1+k (5 x+1) f \sqrt[5]{x}-2 f x^{\frac{6}{5}}\right)}{x^{\frac{6}{5}} (x+1) f} \label{27a} \\ \n\\
\frac{\rho+p}{C}&=&\frac{2  \left(15 c_1 x^2+12 c_1 x+c_1+2 (1-2 k) f x^{\frac{6}{5}}\right)}{x^{\frac{6}{5}} (x+1) f} \label{27b} \\ \n\\
\frac{\rho+3p}{C}&=&\frac{2  \left(3 \left(15 c_1 x^2+12 c_1 x+c_1+2 f x^{\frac{6}{5}}\right)-k \sqrt[5]{x} f (13 x+1)\right)}{x^{\frac{6}{5}} (x+1) f} \label{27c}
\eeq
where we have introduced the substitution $f=(7 x+5)^{33/35}$.

Although undesirable behaviour is evident at the center $x=0$, we investigate, with the aid of  plots, the physical behaviour of this model in the region excluding the center since the core may be filled with a different fluid.  It is evident that  all the physical quantities (electric field intensity, metric potentials, charge density, pressure and the adiabatic speed of sound) are all singular at the origin.  This suggests that our model  may only define a layer within the star excluding the origin.  This means that it represents a spherical shell which contains a non-singular fluid at the core.  For example, we can assume that the core fluid is defined by the Schwarzschild line element.  It is necessary to match the core metric with our singular metric across a common hypersurface.  Then the interior singular shell of the star is matched with the exterior Reissner--N\"{o}rdstrom solution across the boundary interface $r=R$.

\begin{figure}
\begin{center}
\includegraphics[height=4.12 in]{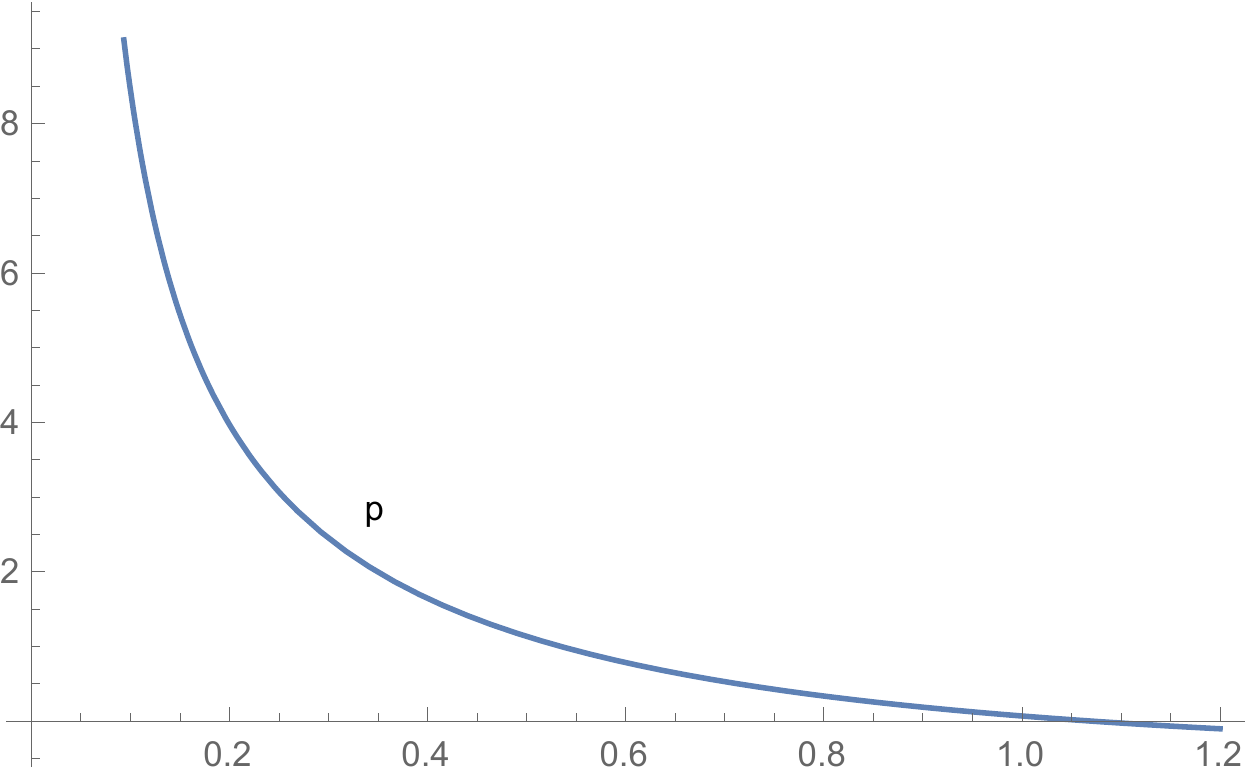}\label{pressure}
\caption{ Pressure $p$ versus the radial component $x$.}
\end{center}
\end{figure}

\begin{figure}
\begin{center}
\includegraphics[height=4.1 in]{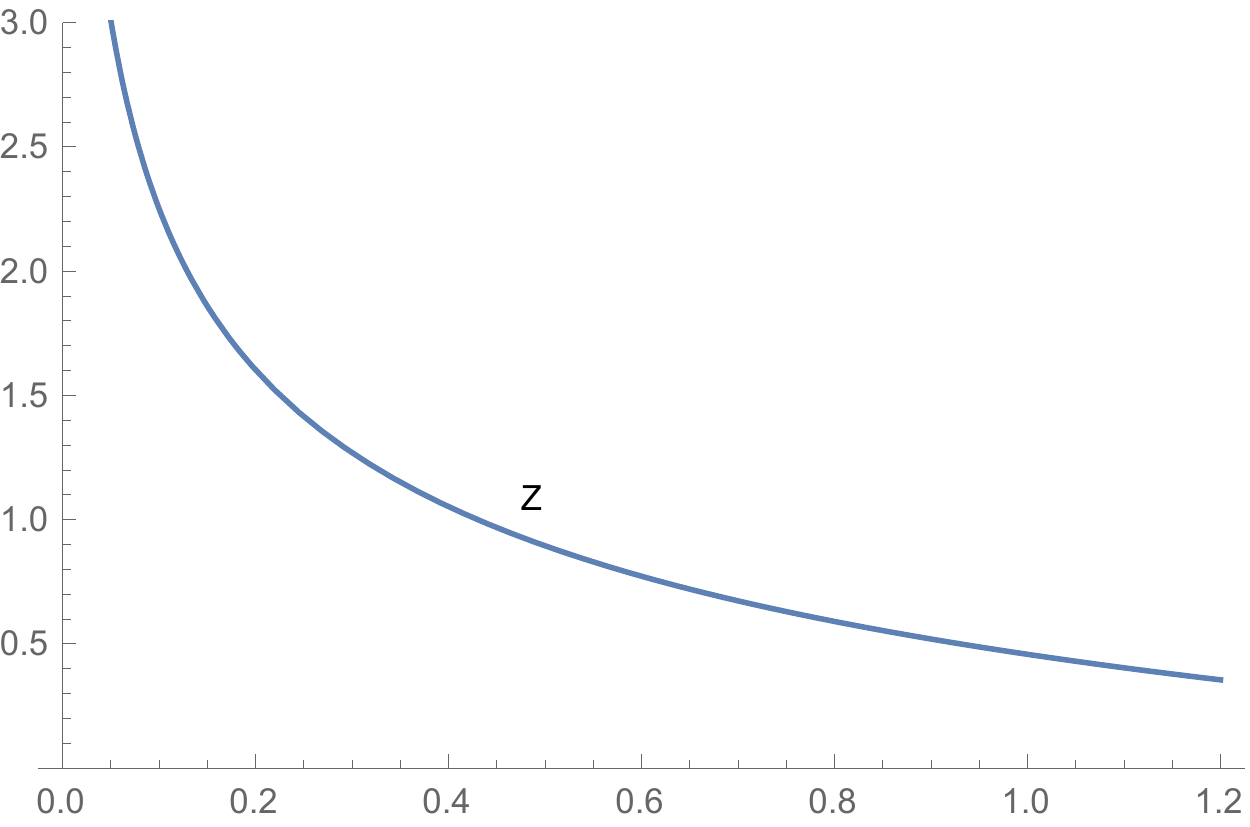}
\caption{Gravitational potential $Z$ versus the radial component $x$.}
\end{center}
\end{figure}

\begin{figure}
\begin{center}
\includegraphics*[height=4.1 in]{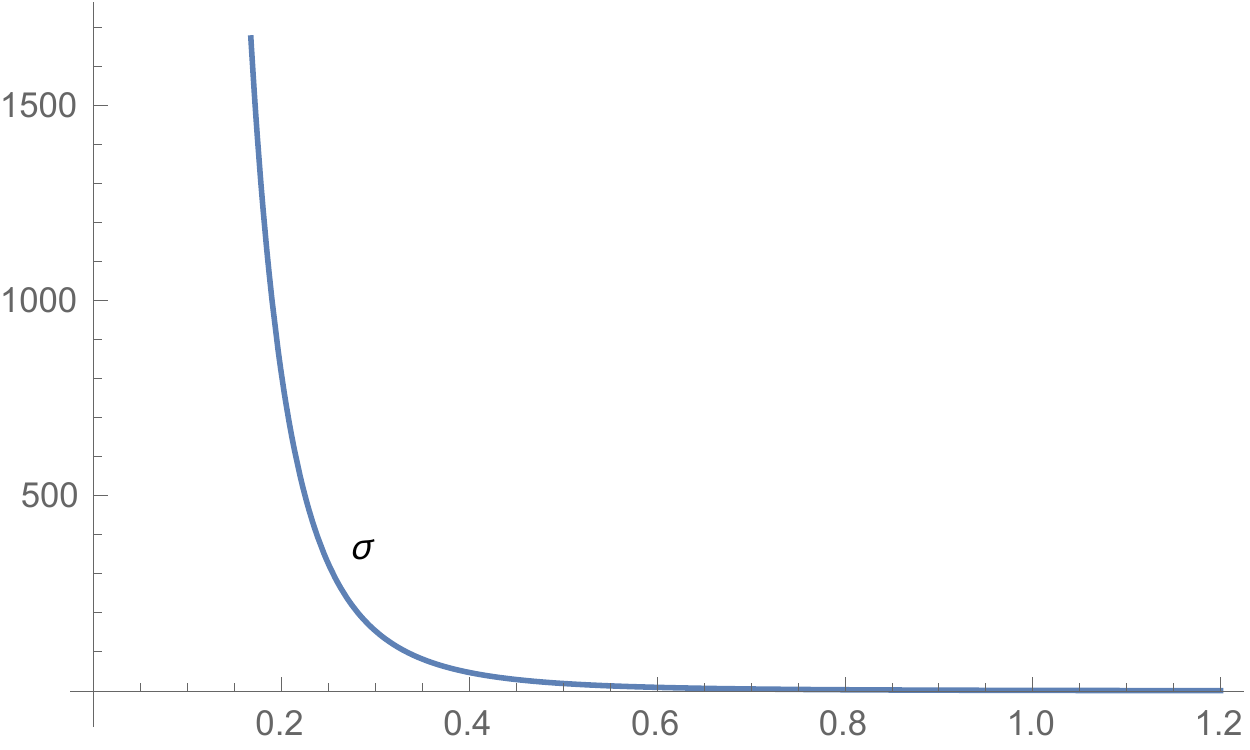}
\caption{Charge density $\sigma$ versus the radial component $x$.}
\end{center}
\end{figure}

\begin{figure}
\begin{center}
\includegraphics*[height=4.1 in]{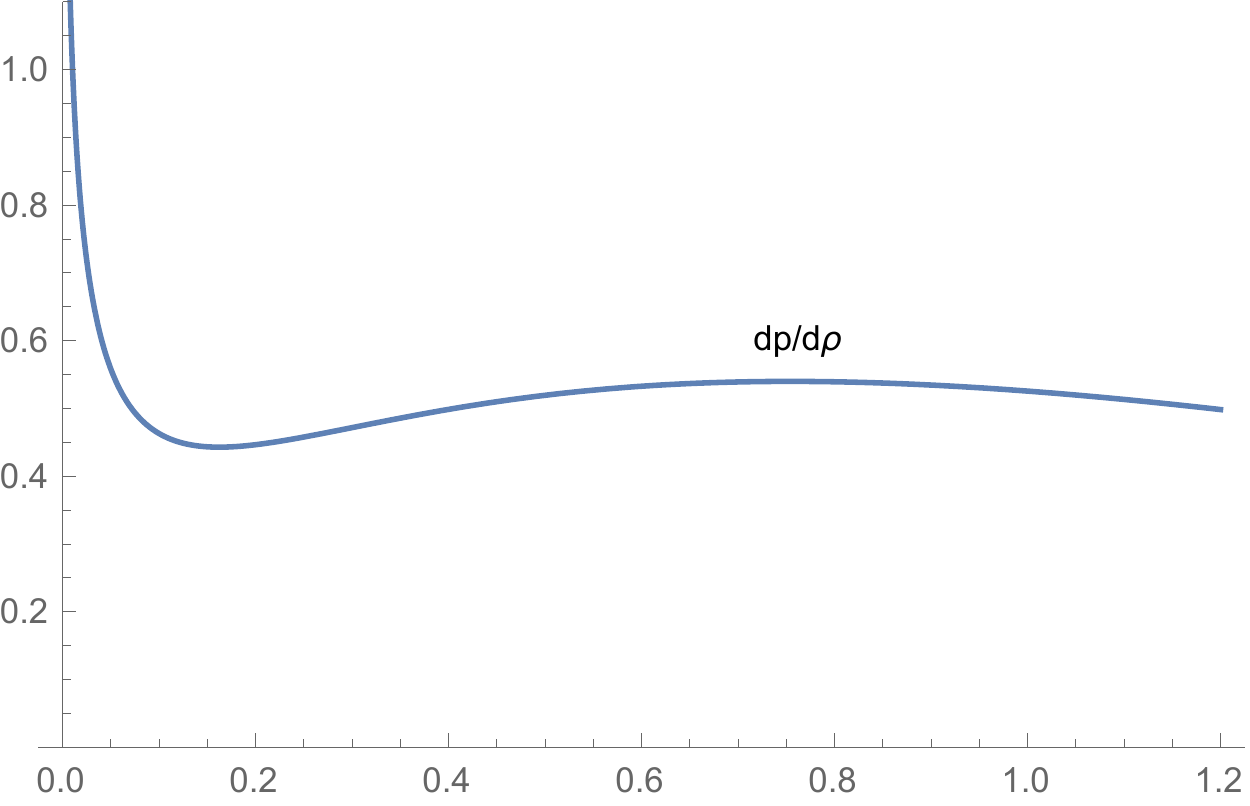}
\caption{Speed of sound  $\frac{dp}{d\rho}$ versus the radial component $x$.}
\end{center}
\end{figure}

\begin{figure}
\begin{center}
\includegraphics*[height=4.1 in]{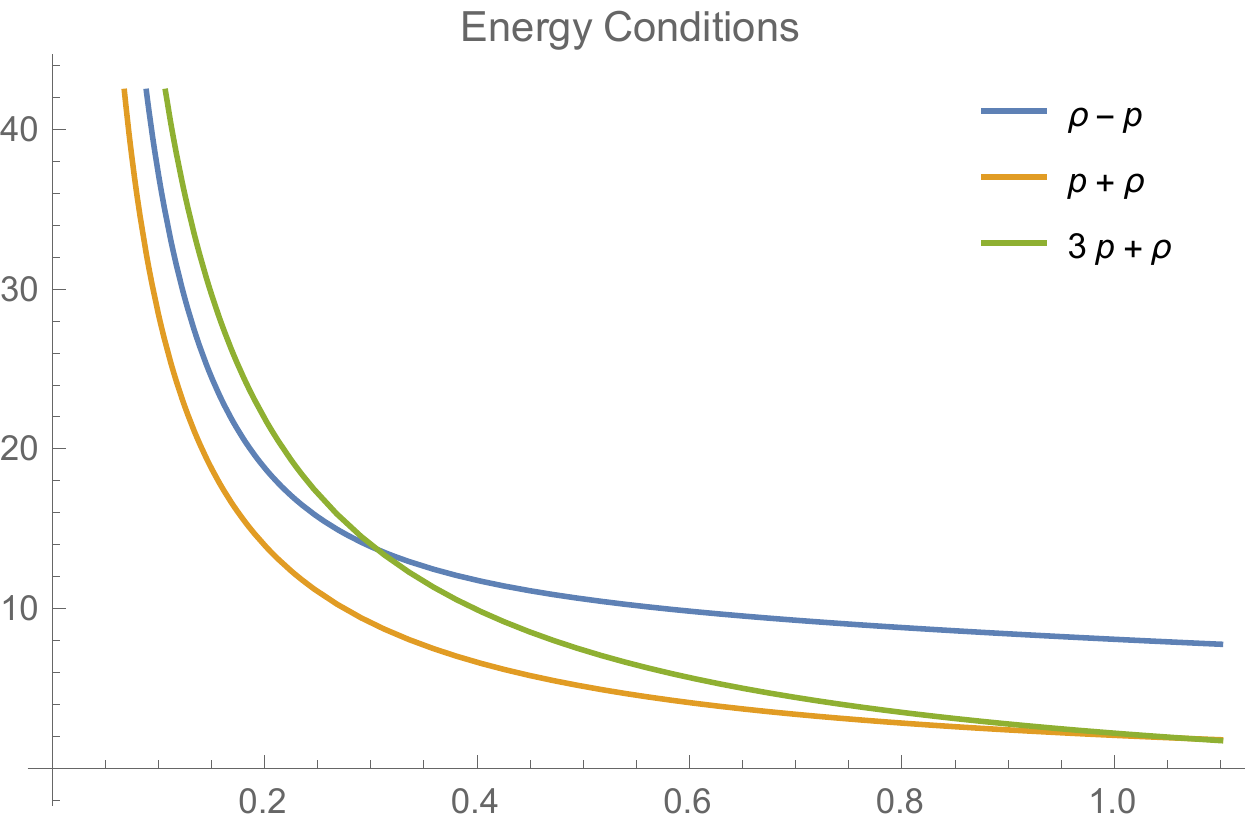}
\caption{Energy conditions versus the radial component $x$.}
\end{center}
\end{figure}

\begin{figure}
\begin{center}
\includegraphics*[height=4.1 in]{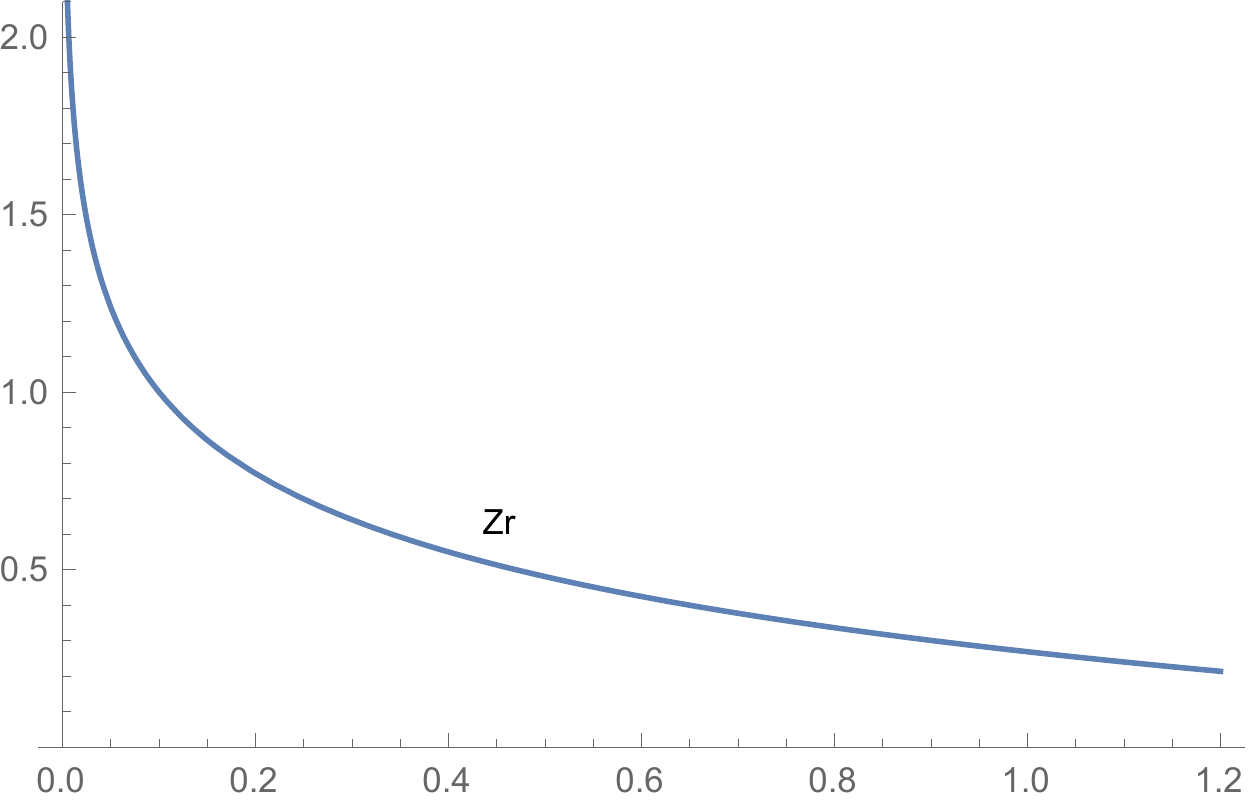}
\caption{Gravitational redshift $z$ versus the radial component $x$.}
\end{center}
\end{figure}

In order to plot these curves the following parametric values have been used: $C=0.5$, $k=1$ and $c_1=3$.  From Figure 1  it is observed that the pressure curve is positive and monotonically decreasing and eventually vanishes at $x=1.15$.  Figure 2 reveals a smooth positive gravitational potential curve.  Figure 3 reflects that the curve of the  charge density  is smooth, positive and monotonically decreasing outwardly.  From Fig 4 it is found that  the curve of the speed of sound  is positive and well within the range [0,1] inside the sphere $x \leq 1.15$ and figure 5 displays the energy conditions curves which are all positive as per the requirements. \\
Taking  the radius of the stellar distribution as $x=1.15$ and using $x=C R^2$ we get $R=1.52$.  The relationship $E^2=\frac{Q^2}{R^4}$
 allows us to obtain the charge--radius ratio as  $\frac{Q^2}{R^2}=1.1777$.   Given that $e^{-2 \lambda}=x+1$ and using the Reissner--Nordstr\"{o}m metric (\ref{532xy}) we find that
$
1-\frac{2M}{R}+\frac{Q^2}{R^2}=C R^2+1
$
and consequently that  $\frac{M}{R}=0.3135$ for the mass--radius ratio.  This ratio satisfies the Buchdahl limit $\frac{M}{R}<\frac{4}{9}$.

We may also compute the gravitational red shift $z$ and this is shown in Fig 6.    Note that that $z<2$ in general within the sphere away from the centre, as expected. These facts suggest that this model of a charged spherical distribution of perfect fluid with inverse square law density decrease, is physically feasible.
The line element for this exact solution is given by
\be
ds^2=-(1+ Cr^2)dt^2  +\left(\frac{c_1 (7 Cr^2 +5)^{\frac{2}{35}}}{\sqrt[5]{Cr^2}}-2 k+1\right)^{-1}dr^2 + r^2 d \Omega^2\label{112qwb2}
\ee

\item{The choice $y=x^{\delta}$}\\

When $y=x^{\delta}$, for any real valued $\delta$, is used in (\ref{28}) we obtain the solution
\begin{equation}
Z= c_1 x^{-\frac{(1-2 \delta)^2}{2 \delta +5}}+\frac{1-2 k}{(1-2 \delta)^2} \label{30a}
\end{equation}
where $c_1$ is an integration constant. All real values of $\delta$ are covered in (\ref{30a}) except $\delta = \frac{1}{2}$. In this case, the solution to the pressure isotropy equation has the form
\be
Z=c_1+\frac{1}{6} (1-2 k) \log (x). \label{31}
\ee
In both cases it is straightforward but tedious to obtain all the remaining dynamical quantities, sound speed and energy conditions. We omit the details in the interest of not being repetitive.

\item{Other forms for $y$}

Various other forms for $y$ such as $y=1+x^n$ and $y=(1+x)^n$ are solvable but only in terms of hypergeometric functions. No other form of $y$ permitted  a complete integration of the field equations in terms of elementary functions.

\end{itemize}

\section{Discussion}

We have investigated the physical viability of a charged isotropic perfect fluid with the isothermal property of an inverse square law fall-off of both density and pressure. The Saslaw {\it{et al}} metric for an isothermal neutral fluid is generalised to include the effects of the electromagnetic field. It was found that the isothermal property was preserved despite the introduction of charge. Then we examined the consequence of dropping the linear equation of state law but retaining the inverse square fall-off of the density. Several classes of exact solutions were developed. Initially we specified functional forms for the spatial gravitational potential and rich classes of solutions expressible as elementary functions emerged. Importantly equations of state, not necessarily linear, were in evidence. When the temporal gravitational potential was prescribed several more new solutions were detected and physically viable equations of state were found. Models were checked for physical plausibility with the aid of plots. It was found that the models studied displayed positive densities and pressures, satisfied the causality criterion as well as the constraints on an acceptable gravitational surface redshift. Moreover, a surface of vanishing pressure existed thus admitting compact or astrophysical objects. The Buchdahl limit on the mass-radius ratio was found to be satisfied and the Bohmer and Harko \cite{boh} lower limit as well as the Andreasson \cite{andr} bound on the mass, charge and radius  were found to be met. Therefore we conclude that whereas the isothermal condition generally yields unbounded cosmological fluids however in the presence of charge bounded distributions emerge.

\bibliography{basename of .bib file}

\end{document}